# Widespread, strong outflows in XQR-30 quasars at the Reionisation epoch

Bischetti* M.[1], Feruglio C.[1,2], D'Odorico V.[1,2,3], Arav N.[4], Bañados E.[5], Becker G.[6], Bosman S. E. I.[5], Carniani S.[3], Cristiani S.[1], Cupani G.[1], Davies, R.[7,8], Eilers A. C.[9], Farina E. P.[10], Ferrara A.[3], Maiolino R.[11], Mazzucchelli C.[12], Mesinger A.[3], Meyer R.[5], Onoue M.[5], Piconcelli E.[13], Ryan-Weber E.[7,8], Schindler J-T.[5], Wang F.[14], Yang J.[15], Zhu Y.[6], and F. Fiore[1,2].

1. INAF - Osservatorio Astronomico di Trieste, Via G. B. Tiepolo 11, I--34143 Trieste, Italy
2. IFPU - Institute for fundamental physics of the Universe, Via Beirut 2, 34014 Trieste, Italy
3. Scuola Normale Superiore, Piazza dei Cavalieri 7, I-56126 Pisa, Italy
4. Department of Physics, Virginia Tech, Blacksburg, VA 24061, USA
5. Max-Planck-Institut für Astronomie, Königstuhl 17, D-69117 Heidelberg, Germany
6. Department of Physics & Astronomy, University of California, Riverside, CA 92521, USA
7. Centre for Astrophysics and Supercomputing, Swinburne University of Technology, Hawthorn, Victoria 3122, Australia
8. ARC Centre of Excellence for All Sky Astrophysics in 3 Dimensions (ASTRO 3D), Australia
9. MIT Kavli Institute for Astrophysics and Space Research, 77 Massachusetts Ave., Cambridge, MA 02139, USA
10. Max Planck Institut für Astrophysik, Karl-Schwarzschild-Straße 1, D-85748 Garching bei München, Germany
11. Kavli Institute for Cosmology, University of Cambridge, Madingley Road, Cambridge, CB3 0HA, UK
12. European Southern Observatory, Alonso de Cordova 3107, Vitacura, Region Metropolitana, Chile
13. INAF - Osservatorio Astronomico di Roma, Via Frascati 33, I– 00078 Monte Porzio Catone, Italy
14. Department of Astronomy, University of Arizona, Tucson, AZ, US
15. Steward Observatory, University of Arizona, Tucson, AZ, US

**Luminous quasars powered by accretion onto billion solar mass black holes already exist at the epoch of Reionisation, when the Universe was 0.5-1 Gyr old[1]. These objects likely reside in over-dense regions of the Universe[2,3], and will grow to form today's giant galaxies. How their huge black holes formed in such short times is debated, particularly as they lie above the local black hole mass - galaxy dynamical mass correlation[4], thus following the *black hole-dominance* growth path[5]. It is unknown what slowed down the black hole growth, leading towards the symbiotic growth observed in the local Universe, and when this process started, although black hole feedback is a likely driver[6]. This deadlock is due to the lack of large, homogeneous samples of high-redshift quasars with high-quality, broad-band spectroscopic information. Here we report results from a *VLT*/X-shooter survey of 30 quasars at redshift $5.8 \lesssim z \lesssim 6.6$ (XQR-30). About 50% of their spectra reveal broad blue-shifted absorption line (BAL) throughs, tracing powerful ionised winds. The BAL fraction in $z \gtrsim 6$ quasars is 2-3 times higher than in quasars at z~2-4.5. XQR-30 BAL quasars exhibit extreme outflow velocities, up to 17% of the light**

**speed, rarely observed at lower redshift. These outflows inject large amounts of energy into the galaxy interstellar medium, which can contrast nuclear gas accretion, slowing down the black-hole growth. The star-formation rate in high-z quasar hosts is generally >100 M$_\odot$/yr[7], so these galaxies are growing at a fast rate. The BAL phase may then mark the beginning of significant feedback, acting first on black hole growth and possibly later on galaxy growth. The red optical colors of BAL quasars at z≳6 indeed suggest that these systems are dusty and may be caught during an initial quenching phase of obscured accretion[8].**

The Ultimate X-shooter Legacy Survey of Quasars at z=5.8-6.6 (XQR-30)[1] has obtained homogeneous, high-quality (median signal-to-noise ratio ≳25 per 50 km/s pixel at 1600-1700 Å), wide-band optical and near-infrared, medium resolution (nominal resolving power 8000-9000 in the full band) spectra of the 30 brightest quasars known at z≳5.8 in the J band (D'Odorico et al. in prep., see Methods for details about the selection of the sample). One of the prime XQR-30 goals is to use this unprecedented sample to study the early growth of supermassive black holes and their feedback on the host galaxies. With a median absolute magnitude $M_{1450Å} = -26.9$ (-27.8 to -26.2), XQR-30 quasars cover the brightest end of the quasar luminosity function at z>5.8 and at any redshift[9].

Black hole driven outflows in quasars can be observed as broad (>2000 km/s) absorption line (BAL) features in the rest-frame UV spectrum[10], blueward of prominent emission lines. To search for these absorption troughs, it is crucial to model the intrinsic rest-frame UV continuum. This is often done by fitting the observed spectra with a continuum model[11]. This method unfortunately has a major drawback: it requires a sufficiently large portion of the spectral band that is free from absorption or emission features, and this condition is usually not met in quasar spectra in the relevant spectral range. We thus adopt a different method to estimate the quasar continuum for the XQR-30 sample. We construct a composite template for every single XQR-30 spectrum by matching the F(1700Å)/F(2100Å) and F(1290Å)/F(1700Å) flux ratios and the equivalent width of the C IV emission line to those of quasars from the SDSS DR7 sample[12,13]. XQR-30 normalised spectra are obtained by dividing each X-shooter spectrum by its matched composite, SDSS template (see Methods and Fig. 3). We systematically searched for absorption troughs associated with the main UV transitions of C IV, Si IV, N V and Mg II ions accessible through the X-shooter spectra. We measured the balnicity index[10], which is a modified equivalent width of the BAL absorption, according to the following definition[11]:

$$BI_0 = \int_0^{v_{lim}} \left(1 - \frac{f(v)}{0.9}\right) C dv \quad (1)$$

where $f(v)$ is the normalised spectrum, $C = 1$ if $f(v) < 0.9$ for contiguous troughs of >2000 km/s, $C = 0$ otherwise, and $v_{lim} = 64500$ km/s (see Methods). Traditionally, BALs are defined to have $BI_0$>0 km/s. However, the uncertainties on $BI_0$, dominated by the accuracy of the composite template, allows us to measure with high reliability $BI_0 \gtrsim 100$ km/s in our XQR30 spectra and in a reanalysis of SDSS spectra of quasars at lower redshift, see below. We identify the minimum $v_{min}$ and maximum $v_{max}$ BAL velocity for a given transition as the lowest and highest velocity for which $C = 1$ in Eq. (1), respectively. Our approach differs from most identification studies of BAL quasars in SDSS, that typically adopt $v_{lim} = 25000$ km/s[11,14,15]. The latter choice accounts only for BAL features in the spectral region between C IV and Si IV and, therefore, underestimates $BI_0$ for BAL systems with $v_{max} > 25000$ km/s, and completely misses BAL systems with $v_{min} > 25000$ km/s. We identify 14 XQR-30 BAL quasars (all showing C IV BAL, 8 showing also a Si IV BAL, 3 a N V BAL and 1 a Mg II BAL). The BAL fraction in the XQR-30 sample is thus $47^{+16}_{-12}$%, considering 90% confidence level for a Poisson distribution. By conservatively excluding 3 quasars at z~6.0, in which the weak BAL absorption falls close to edge of the X-Shooter VIS arm and, therefore, might be affected by a low X-Shooter response curve (see Methods), the XQR-30 BAL fraction remains as high as $41^{+16}_{-12}$%. Recent works collecting large spectroscopic samples of $z \gtrsim 5.7$ quasars (some of which are in common with XQR-30) have reported a BAL fraction of ~16%[16,17], lower than the results from XQR-30. However, this value has to be considered as a lower limit on the actual BAL fraction, resting on visual inspection only, sometimes being limited to a small range of $v_{lim}$ ($\lesssim 10000$ km/s) and based on spectra whose resolution and/or signal-to-noise ratio are significantly lower than in XQR-30. Previous hints for a BAL quasar fraction as high as 50% at $z \gtrsim 5.7$ were based on very small quasar samples[18]. Note that these are the "observed" BAL fractions. In the following we compare these BAL fractions to those estimated in quasar samples at different redshift and with matched selection criteria (see Methods).

---
[1] PID 1103-A.0817, P.I. V. D'Odorico, 248h with X-shooter at the ESO Very Large Telescope

A BAL fraction of 40-50% is significantly higher than the BAL fraction of 10-15% typically observed in z~2-4 SDSS quasars[11,15,19]. A more quantitative comparison with lower redshift quasar samples requires the application to control samples of selection criteria and BAL identification method similar to that used for the XQR-30 sample, because some of the mismatch with the literature result could be due to different assumptions in these tasks. We built and analysed a number of control samples by using the same selection criteria and analysis followed for the XQR-30 sample. We first selected from SDSS a quasar sample at z=2.1-3.2 requiring a detection in the 2MASS H and K bands, that is matching the XQR-30 rest-frame selection criteria (see Methods), with median $M_{1450Å} = -26.5$. We then searched for BALs in this SDSS control sample using the same BAL identification method used for XQR-30 quasars. We found 307 BAL systems out of 1580 quasars, that is a BAL fraction of $19.4 \pm 1.1$% ($BI_0 >$ 100 km/s), 2.5 times lower that the BAL fraction at z≃5.8-6.6. We calculate the probability that the BAL fraction in XQR-30 traces the same BAL quasar population of SDSS z~2-3 quasars by generating $10^6$ subsamples of 30 randomly selected, non-repeated quasars from the SDSS comparison sample, and computed the probability to observe ≥14 BALs out of 30 quasars as $p_{rand}(BI_0) = N_{\geq 14}/10^6$, where $N_{\geq 14}$ is the number of subsamples with ≥14 BALs. This probability is $p_{rand}(BI_0) = 2 \times 10^{-4}$. We also performed a Mann-Whitney U test[20] for the hypothesis that the $BI_0$ distributions of XQR-30 and the SDSS control sample in Figure 1 (top left) are equal. This hypothesis is rejected with a probability $p_U(BI_0) = 2.6 \times 10^{-4}$, similar to the previous value. If we exclude the three quasars whose BAL absorption falls close to the edge of the X-Shooter VIS arm, we obtain $p_{rand}(BI_0) = N_{\geq 11}/10^6 = 3.3 \times 10^{-3}$ and $p_U(BI_0) = 2.8 \times 10^{-3}$.

We verified that the BAL fraction in the control SDSS sample does not significantly vary with the S/N of the spectra, consistent with previous studies of SDSS BAL quasars[11]. We investigated whether the higher BAL fraction observed in XQR-30 quasars with respect to SDSS control sample quasars is due to different nuclear properties. Figure 2 shows the Eddington ratio $\lambda_{Edd} = L_{bol}/L_{Edd}$, where $L_{bol}$ and $L_{Edd}$ are the quasar bolometric and Eddigton luminosities, as a function of the black hole mass ($M_{BH}$) for the XQR-30 quasars and the SDSS control sample quasars. Figure 2 shows that XQR-30 quasars lie at the upper envelope of the $\lambda_{Edd}$ distribution of the SDSS control sample. We then calculated the BAL fraction in a subset of SDSS quasars from the control sample with $\lambda_{Edd}$ and $M_{BH}$ matching the XQR-30 sample (Figure 2). This gave 19±3%, consistent with the BAL fraction of the full SDSS control sample quasars and again about 2.5 times smaller than the BAL fraction in XQR-30 quasars. Similarly, the SDSS BAL fraction does not increase when selecting only the brightest sources, probing the same quasar luminosity range of the XQR-30 sample ($L_{bol} \gtrsim 10^{47}$ erg/s). Finally, as the redshift range 2.13<z<3.20 contains redshift intervals with low selection completeness of SDSS quasars, we also consider two SDSS quasar control samples in the redshift intervals 2.13<z<2.4 and 3.6<z<4.6 which are less affected by completeness issues, concluding that selection completeness does not significantly affect the BAL fraction measured using Eq. 1 (see Methods).

BAL XQR-30 quasars with the most powerful outflows ($BI_0 > 1000$ km/s) show redder rest-frame optical colors than non-BAL quasars, as traced by *WISE*[21] W1-W2≃0.1-0.4 color (Figure 1 top right), while the total W1-W2 color distribution of XQR-30 quasars reflects z≳6 quasars mid-infrared selection criteria[22,23,24]. The lack of XQR-30 quasars with very red optical colors (W1-W2>0.4) is likely related to the UV sample selections, introducing a bias against dust-obscured sources. Redder W1-W2 colors for BAL XQR-30 quasars suggest a link between the slope of the optical spectrum and the presence of strong BAL outflows at z≳6. BAL XQR-30 quasars may possibly be dustier than non-BAL quasars. *WISE* photometry does not allow us to assess whether this dust is located close to the nucleus (e.g. in the torus or in the quasar broad/narrow line region), it is embedded in the BAL clouds, or it extends on galaxy scale. In the SDSS control sample, no difference in the rest-frame optical colors of BAL and non-BAL quasars, as traced by 2MASS H-K color[12,25], is observed.

Figure 1 compares the $v_{min}$ and $v_{max}$ distributions of the XRQ-30 quasars with those of the SDSS comparison sample. More than half of the XQR-30 C IV BAL quasars have $v_{min} > 15000$km/s, and all but two have $v_{max} > 20000$km/s, velocities hardly ever observed in the SDSS control sample quasars, which indeed show median $<v_{min}> = 3700$km/s and $<v_{max}> = 14000$km/s. Following the same approach described above for the BAL fraction, we have calculated the probability of observing by chance in the SDSS BAL control sample 57% of the BAL quasars with $v_{min} > 15000$km/s and 86% of the BAL quasars with $v_{max} > 20000$km/s. These probabilities are $p_{rand}(v_{min}) = 4 \times 10^{-5}$ and $p_{rand}(v_{max}) < 10^{-6}$. The U test for the hypothesis that the XQR30 and SDSS control sample $v_{max}$ distributions are drawn from the same parent population gives a probability $p_U(v_{max}) = 8 \times 10^{-5}$. We also note that 5 XQR-30 quasars (17%) show extremely high velocity BAL outflows ($v_{max} >$ 30000 km/s), while these objects have been found to represent only ~0.5% of the total SDSS quasar population at z≲4.5[26]. Two extremely high velocity BAL quasars have been recently identified also at z≳7[27,1]. We conclude that BAL winds in z>5.8 quasars are significantly faster than at lower redshift. The extreme observed velocities might be explained by the presence of dust mixed with the BAL clouds, because of the higher radiation boost efficiency on dust than on the ionised gas[28,29]. A redshift evolution of the kinematic properties of black-hole driven outflows, and higher outflow velocities in z≳6 quasars have been also suggested by recent studies of UV emission line blueshifts[17].

X-shooter data alone do not allow us to accurately evaluate the outflowing wind mass (and thus the wind mass outflow rate and the kinetic energy rate) for the XQR-30 BAL sample, since this generally requires the measurement of non-saturated absorption lines such as SIV/SIV* 1062.7,1072.9 Å[30,31], which are embedded in a dense Lyα forest at z~6, and thus impossible to study at these high redshifts. Should the wind masses at z>5.8 be similar to those of lower redshift BALs with mass measurements[32,33], the XQR-30 quasar kinetic power ($\dot{E}_{kin}$) would be >10 times higher than in lower redshift BAL quasars, because of the systematically higher $v_{max}$. Since $\dot{E}_{kin}$ of BAL winds in low-z quasars is typically in the range 0.001-0.03$L_{bol}$[34], $\dot{E}_{kin}$ of XQR-30 BAL quasars is likely 0.01-0.3$L_{bol}$, a huge kinetic power injected into the host galaxies. Furthermore, the high BAL fraction at z≳6 strongly points toward a scenario in which either the wind geometry or the timescales of the AGN wind phases evolve with redshift. Either wider angle winds or longer wind blow-out phases translate into higher BAL fractions. Specifically, the BAL fraction measured in XQR-30 quasars corresponds to a BAL divergence angle of 20-30º, considering the standard scenario for the BAL structure[35]. It also implies that the BAL outflow timescale is about half the duration of the active phase of the supermassive black hole powering the quasar. In both cases, a large BAL fraction implies high total (integrated over the active galactic nucleus timescale) energy deposited by the wind in the interstellar medium. With respect to quasars at z~2-3, quasars in the early Universe (z≳5.8) collectively inject in their hosts a factor of 20-40 times more energy (a factor of ~10 because of their higher velocities and a factor of ~2.5 for the higher BAL fraction). This energy injection can likely inhibit further gas accretion, slowing down the paroxysmal BH growth. At the same time, high-z quasars hosts are growing quite rapidly, since star-formation rates up to thousands M⊙/yr are typically observed[36,37], even assuming that a large fraction of the host galaxy dust is heated by the AGN, as found in lower redshift quasars of similar bolometric luminosity[38]. Therefore, high-z BAL quasars may be experiencing the onset of significant AGN feedback, marking the transition from *BH dominance* to *BH-galaxy symbiotic* growth phases. Cosmological, hydrodynamic simulations of early black hole and galaxy evolution support this scenario by identifying z~6-7 as the transition epoch during which AGN feedback increases in strength and starts to significantly slow down the black hole growth[3,6], exponential at z>7, while the host galaxy is still rapidly growing.

ALMA observations of XQR-30 BAL quasars will be fundamental to compare either the dust and SFR properties with those of non-BAL quasars at the Reionisation epoch. A difference in the dust properties will favour a scenario of BAL quasars being caught in a peculiar evolutionary phase, while no distinction will imply a redshift evolution in terms of the BAL outflow geometry. Finally, a measure of the SFR will allow us to determine whether $z \gtrsim 6$ BAL quasars are experiencing a different growth phase than non-BAL quasars and if BAL quasars can be used as tracers of ongoing black hole feedback at these early epochs.

**The XQR-30 sample properties and BAL system parameters**

| Name | Ra | Dec | SNR | $z_{Mg\,II}$ | J | W1 | W2 | $BI_0$ | $v_{min}$ | $v_{max}$ | Type | Ref. |
|---|---|---|---|---|---|---|---|---|---|---|---|---|
| (1) | (2) | (3) | (4) | (5) | (6) | (7) | (8) | (9) | (10) | (11) | (12) | (13) |
| PSOJ007+04 | 00 28 06.56 | +04 57 25.64 | 79 | 5.954 | 19.77 | 19.97 | 19.75 | - | - | - | no-BAL | 22,22 |
| PSOJ009−10 | 00 38 56.52 | −10 25 53.90 | 27 | 5.938 | 19.93 | 19.16 | 19.00 | 1110 | 33040 | 38380 | BAL | 22,39 |
| PSOJ023−02 | 01 32 01.70 | −02 16 03.11 | 28 | 5.816 | 19.78 | 19.20 | 18.82 | 2120 | 3950 | 21240 | BAL | 22,22 |
| PSOJ025−11 | 01 40 57.03 | −11 40 59.48 | 32 | 5.816 | 19.65 | 19.37 | 19.26 | - | - | - | no-BAL | 22,22 |
| PSOJ029−29 | 01 58 04.14 | −29 05 19.25 | 29 | 5.976 | 19.07 | 18.81 | 18.58 | - | - | - | no-BAL | 22,22 |
| ATLASJ029−36 | 01 59 57.97 | −36 33 56.6 | 30 | 6.013 | 19.57 | 19.28 | 19.19 | - | - | - | no-BAL | (*),39 |
| VDESJ0224−4711 | 02 24 26.54 | −47 11 29.4 | 38 | 6.525 | 19.73 | 18.78 | 18.70 | - | - | - | no-BAL | 39,39 |
| PSOJ060+24 | 04 02 12.69 | +24 51 24.42 | 34 | 6.170 | 19.71 | 19.17 | 19.41 | - | - | - | no-BAL | 22,22 |
| J0408−5632 | 04 08 19.23 | −56 32 28.8 | 106 | 6.033 | 19.85 | 20.04 | 19.85 | 360 | 24130 | 28820 | BAL | 39,39 |
| PSOJ065−26 | 04 21 38.05 | −26 57 15.60 | 135 | 6.179 | 19.36 | 19.01 | 19.00 | - | - | - | no-BAL | (*),22 |
| PSOJ065+01 | 04 23 50.15 | +01 43 24.79 | 25 | 5.804 | 19.74 | - | - | 920 | 16890 | 38610 | BAL | (*) |
| PSOJ089−15 | 05 59 45.47 | −15 35 00.20 | 102 | 5.972 | 19.17 | 18.18 | 17.80 | 13910 | 2250 | 30360 | BAL | 22,39 |
| PSOJ108+08 | 07 13 46.31 | +08 55 32.65 | 126 | 5.955 | 19.07 | 18.69 | 18.51 | - | - | - | no-BAL | 39,22 |
| SDSSJ0842+1218 | 08 42 29.43 | +12 18 50.58 | 133 | 6.067 | 19.78 | 19.00 | 19.03 | 740 | 19920 | 27380 | BAL | 40,39 |
| J0923+0402 | 09 23 47.12 | +04 02 54.4 | 30 | 6.626 | 20.28 | 19.20 | 19.05 | 13570 | 6090 | 29500 | BAL | 41,39 |
| PSOJ158−14 | 10 34 46.50 | −14 25 15.58 | 30 | 6.065 | 19.19 | 18.62 | 18.47 | - | - | - | no-BAL | 42,39 |
| PSOJ183+05 | 12 12 26.98 | +05 05 33.49 | 113 | 6.428 | 19.77 | 19.74 | 20.03 | - | - | - | no-BAL | 43,39 |
| PSOJ183−12 | 12 13 11.81 | −12 46 03.45 | 30 | 5.893 | 19.07 | 18.98 | 19.18 | 540 | 16280 | 25910 | BAL | 39,22 |
| PSOJ217−16 | 14 28 21.39 | −16 02 43.30 | 34 | 6.135 | 19.71 | 18.99 | 19.39 | - | - | - | no-BAL | 39,22 |
| PSOJ217−07 | 14 31 40.45 | −07 24 43.30 | 113 | 6.166 | 19.85 | 19.92 | 19.67 | 4410 | 22000 | 43970 | BAL | 39,39 |
| PSOJ231−20 | 15 26 37.84 | −20 50 00.66 | 37 | 6.564 | 19.66 | 19.91 | 19.97 | 600 | 540 | 2680 | BAL | 44,39 |
| J1535+1943 | 15 35 32.87 | +19 43 20.1 | 36 | 6.358 | 19.63 | 18.55 | 18.46 | - | - | - | no-BAL | 39,39 |
| PSOJ239−07 | 15 58 50.99 | −07 24 09.59 | 164 | 6.114 | 19.37 | 18.94 | 18.65 | 1890 | 810 | 6100 | BAL | 42,39 |
| PSOJ242−12 | 16 09 45.53 | −12 58 54.11 | 15 | 5.840 | 19.71 | 19.00 | 19.50 | - | - | - | no-BAL | 39,22 |
| PSOJ308−27 | 20 33 55.91 | −27 38 54.60 | 85 | 5.799 | 19.46 | 19.63 | 19.37 | - | - | - | no-BAL | 22,22 |
| PSOJ323+12 | 21 32 33.19 | +12 17 55.26 | 103 | 6.585 | 19.74 | 19.06 | 18.97 | - | - | - | no-BAL | 43,22 |
| VDESJ2211−3206 | 22 11 12.17 | −32 06 12.94 | 74 | 6.330 | 19.54 | 18.92 | 18.80 | 6840 | 9240 | 24340 | BAL | (*),22 |
| VDESJ2250−5015 | 22 50 02.01 | −50 15 42.2 | 96 | 5.985 | 19.17 | 19.11 | 19.04 | 6330 | 22500 | 50320 | BAL | 39,22 |
| SDSSJ2310+1855 | 23 10 38.89 | +18 55 19.70 | 161 | 5.992 | 18.86 | 18.50 | 18.75 | 590 | 18520 | 26890 | BAL | 39,22 |
| PSOJ359−06 | 23 56 32.45 | −06 22 59.26 | 154 | 6.169 | 19.85 | 19.26 | 19.04 | - | - | - | no-BAL | 22,22 |

**Table 1.** 1) quasar name, (2-3) coordinates, (4) median signal to noise per pixel in the 1600-1700 Å wavelength range, for a 50 km/s spectral resolution. In the case of PSOJ242-12, two hours of observations are still needed to obtain the final spectrum; (5) Mg II-based redshift, (6) AB magnitude in the J band, (7-8) AB magnitude in the W1 and W2 bands, (9-11) balnicity index, minimum and maximum velocity of the C IV BAL outflows, in units of km/s. Strong(weak) BAL outflows typically have $BI_0 > 1000$ km/s($BI_0 < 1000$ km/s). Positive $v_{min}$ and $v_{max}$ values indicate blue-shifted C IV absorption. (12) BAL/no-BAL classification. (13) reference for the J and *WISE* magnitudes, respectively. J magnitudes indicated with (*) are presented in this work (Methods).

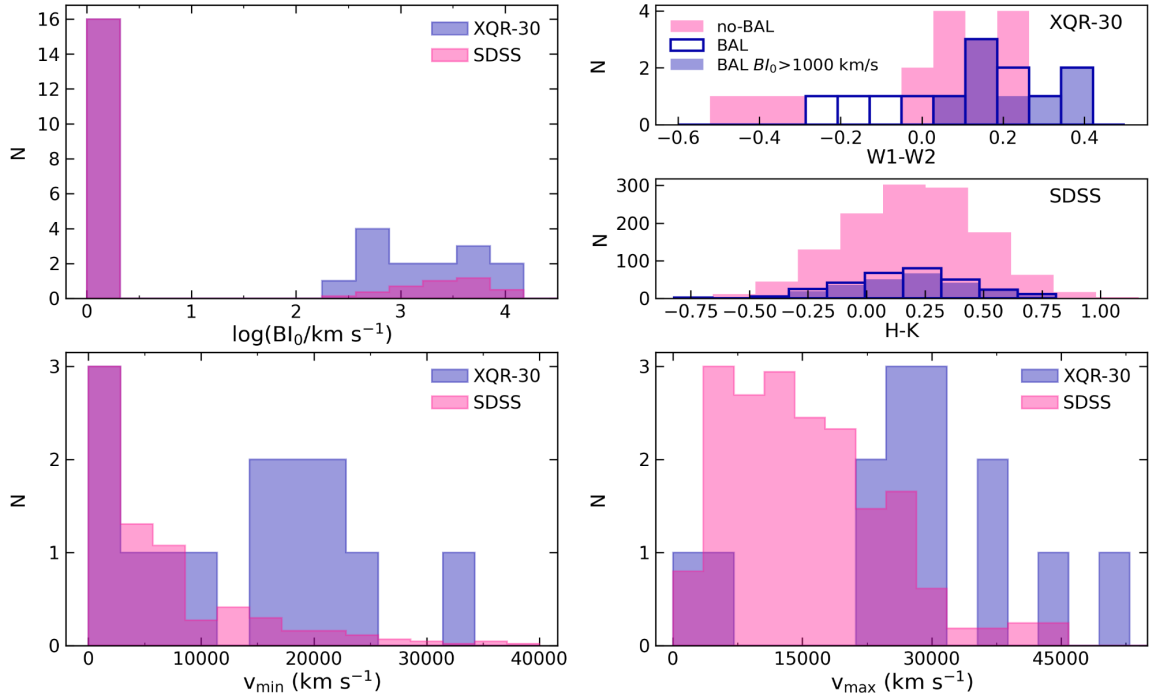

**Figure 1.** Properties of C IV BAL quasars. [Top left panel] The total (including BAL and non-BAL quasars) balnicity index distribution for the XQR-30 sample, compared with that of SDSS quasars at z~2-3. Non-BAL quasars are shown as $\log(BI_0/\mathrm{km\,s^{-1}}) = 0$. The SDSS distribution has been rescaled to the maximum of the XQR-30 distribution. [Top right panels] Rest-frame optical color distribution, as traced by W1-W2 color for XQR-30 quasars, and the equivalent H-K colour distribution for the control SDSS sample. All magnitudes are in the AB system. In total 13 XQR-30 BAL quasars are shown, because the *WISE* photometry of PSOJ065+01 is contaminated by nearby sources. [Bottom panels] Minimum(left) and maximum(right) velocity distributions associated with the C IV absorption troughs in BAL quasars.

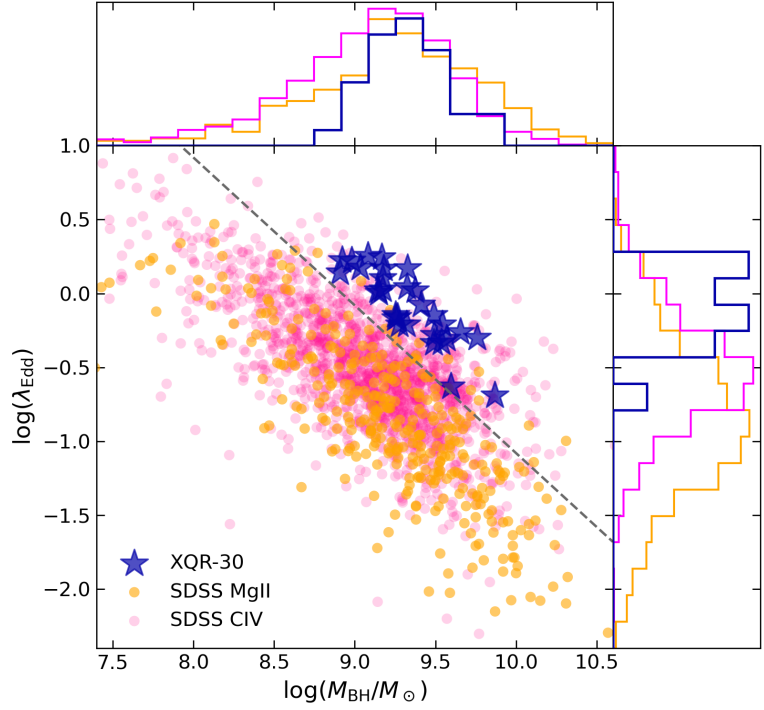

**Figure 2.** Eddington rate $\lambda_{Edd}$ as a function of black-hole mass $M_{BH}$. XQR-30 quasars are shown as blue stars while SDSS control sample quasars are indicated by orange and magenta circles. Top(right) histogram shows the $M_{BH}(\lambda_{Edd})$ distributions for the two samples. $M_{BH}$ of XQR-30 quasars and SDSS quasars with z<2.3 have been derived from the Mg II line, while for z>2.3 SDSS quasars it is based on the C IV line, correcting for non-virial motions (see Methods). To build a $M_{BH}$ and $\lambda_{Edd}$ matched sample of SDSS quasars, sources above the dashed line in the main panel have been considered.

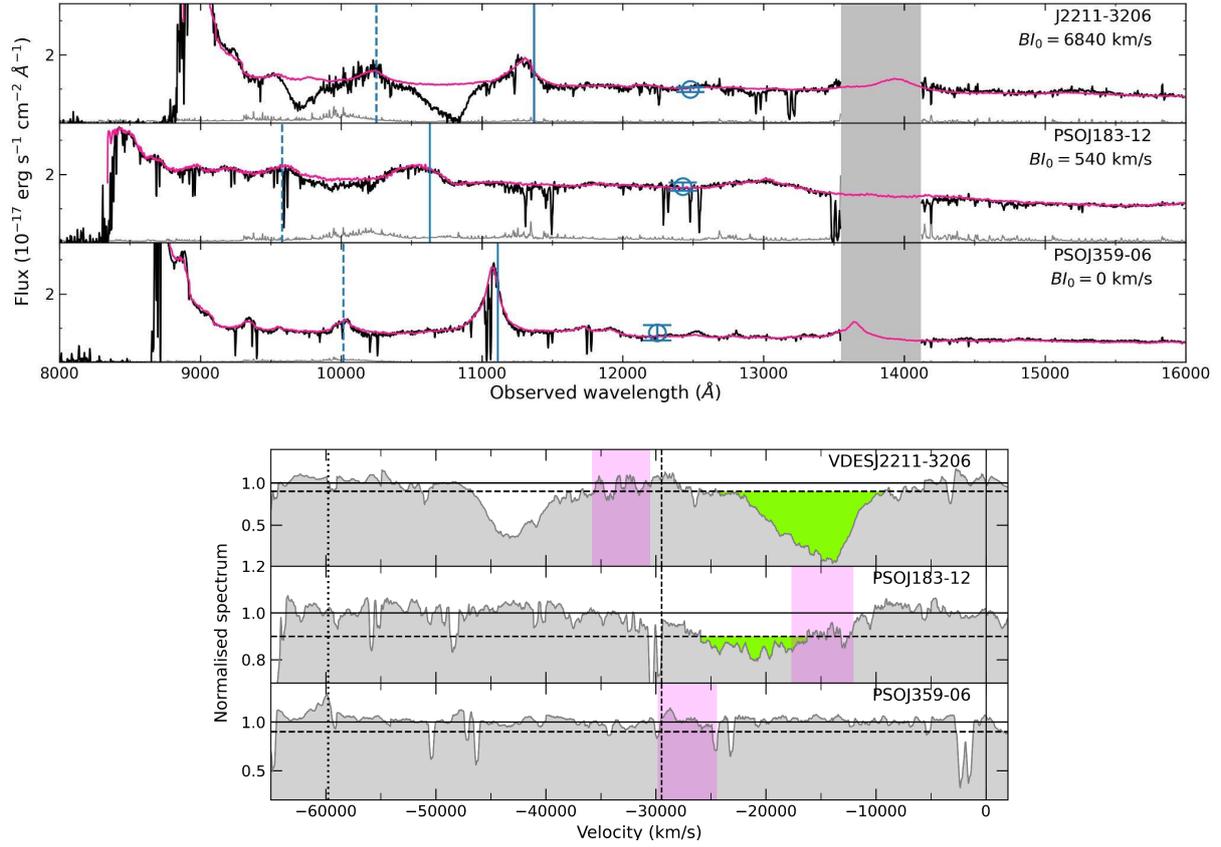

**Figure 3.** [Top panels] Examples of X-shooter spectra of XQR-30 quasars showing strong ($BI_0$>1000 km/s), weak ($BI_0$<1000 km/s) and no ($BI_0$=0 km/s) BAL absorption features. Spectra have been binned in three pixels and the flux uncertainty multiplied by a factor of five is shown in gray. The composite template, used to estimate the intrinsic quasar emission, is indicated by the magenta curve. Vertical solid(dashed) line corresponds to the position of the C IV(Si IV) emission line according to $z_{Mg\,II}$. The grey shaded area identifies the spectral window affected by strong telluric absorption. Blue circles represent J-band magnitudes. [Bottom panels] Corresponding normalised spectra, smoothed to 500 km/s. The velocity axis in each panel is relative to the rest-frame wavelength of C IV. Vertical solid, dashed, and dotted lines indicate the position of C IV, Si IV and N V, respectively. The magenta area highlights the overlapping spectral region between the X-shooter VIS and NIR arms in which the uncertainty on the X-shooter response curve is larger (see Methods). BAL systems are highlighted as green shaded areas.

**Methods**

**XQR-30 sample and X-shooter observations.** XQR-30 quasars were selected to match the following requirements: declination δ< +27 deg, to be observable from ESO/*VLT*; redshift z ≥ 5.8, to require the Mg II emission line in the K-band; J magnitude $J_{AB}$≤19.8 for z<6.0 sources and $J_{AB}$≤20.0 for quasars at z≥6.0; no existing deep X-shooter data, defined as S/N>25 per pixel. The application of those constraints to the literature[22,45] and newly discovered quasars, resulted in a sample of 30 sources covering the redshift range 5.8≤z≤6.6 (Table 1). The X-shooter spectra have been acquired with slit widths of 0.9 arcsec and 0.6 arcsec, and resolving power of R~8900 and 8100, in the visible (VIS) and near-infrared (NIR) arms, respectively. The observing time on target ranges from 4h to 11h. The median SNR per pixel in the rest-frame 1600-1700 Å wavelength range is between 25 and 160 for spectra rebinned to 50 km/s, as used in this work. Raw frames have been reduced with a custom software pipeline developed within the collaboration[46]. A relative flux calibration has been applied to all reduced spectra by considering the instrument response curve built from a single spectro-photometric standard star for all frames. A comparison between the spectra obtained by this method with those reduced with the ESO pipeline[47] and calibrated with standard stars acquired the same night of the observation, does not show significant differences. X-shooter VIS and NIR spectra have been combined with the Astrocook software[48] by scaling the NIR to the VIS using the median values computed in the overlapping spectral region 10000-10200 Å (in the observed frame). We checked that this procedure had not introduced any spectral artifact in the X-shooter data by visual inspection of all XQR-30 spectra. We note that the ~10000 Å spectral region (hatched area in Fig. 3, Fig. S1 and Fig. S2) can be spuriously affected by an anomalous drop in the X-shooter response curve of the VIS arm. If this effect is not properly corrected, it may result in a flux drop in the X-shooter spectrum mimicking weak BAL absorption between C IV and Si IV for a *z*~6.0 quasar. We have verified that this effect does not evidently affect BAL quasars in XQR-30 because no significant variation in the normalisation of the VIS spectrum around 10000 Å is observed between different exposures. Nevertheless, as a minor flux drop not detectable in the individual exposures might still be present, we also report the BAL fraction in XQR-30 after excluding the three weak BAL quasars (namely J0408-5632, SDSSJ0842+1218, SDSSJ2310+1855) in which the absorption falls in the overlap region. The absolute flux calibration has been obtained by normalising the combined spectrum to the J band photometry.

**Imaging observations.** Four of the quasars in the sample had no published J-band information in the literature and we present their J-band AB magnitudes in Table 1, namely ATLASJ029-36, PSOJ065+01, PSOJ065-26, and J2211-3206. For J2211-3206, we report the magnitude from the VIKING DR4 catalog[49]. We observed ATLASJ029-35 and PSOJ065-26 on 24 September 2017 for 6 minutes each with the Fourstar infrared camera[50] at the Magellan Baade 6.5m telescope at Las Campanas observatory. PSOJ065+01 was observed on 21 December 2018 for 5 minutes with the SofI camera[51] mounted on the New Technology Telescope at La Silla observatory. The images were processed with standard reduction steps (bias subtraction, flat fielding, sky subtraction, stacking) and the photometry was calibrated against stellar sources in the 2MASS catalog.

**SDSS Control samples.** We build a first control sample by matching the selection criteria of the XQR-30 sample. All XQR-30 quasars have been discovered via rest-frame UV colors probing the Lyα break. An additional *WISE*[22] W1(3.4μm)-W2(4.6μm) color selection can be used to discriminate between z≳6 moderately-luminous quasar candidates and contaminants in the photometric searches[22,23,24,43], which requires these quasars to be detected by *WISE* in rest-frame optical. Starting from a catalogue of SDSS DR7 quasars[12,13], we selected the 11800 quasars in the redshift range 2.13<z<3.20 for which the 1216-2100Å wavelength range is covered by SDSS spectroscopy and the H(1.7μm) and K(2.2μm) bands probe similar rest-frame spectral regions covered by W1 and W2 bands for the XQR-30 quasars. As a rest-frame optical selection has been found to increase the BAL fraction in z~2-3 quasars[52], we conservatively require the control sample to be detected by the 2MASS survey[25] in both H and K bands. We exclude noisy spectra by requiring a median S/N>5 in the 1500-1600Å range. The resulting sample consists of 1580 quasars, for which we measure a BAL fraction of 19.4 ± 1.1% ($BI_0$> 100 km/s). We also study two subsamples of the first control sample with: (i) high $\lambda_{Edd}$= $L_{bol}/L_{Edd}$ and 2.13<z<3.20, where $L_{bol}$ is the quasar bolometric luminosity, and $L_{Edd}$ is the Eddington luminosity, including 268 quasars above the black dashed line in Fig. 2. For this subsample, matching the $M_{BH}$ and $\lambda_{Edd}$ distributions of XQR-30 quasars, the BAL fraction is 19.0±3.0%; (ii) $L_{bol}$≳$10^{47}$ erg/s and 2.13<z<3.20, that is probing the same luminosity range of the XQR-30 sample (275 quasars, BAL fraction 18.9±3.0%). The same BAL fraction is recovered when requiring a match in absolute magnitude $M_{1450Å}$.

As the redshift range 2.13<z<3.20 contains redshift intervals with low selection completeness[53,19], we also consider a subsample with (iii) 2.13<z<2.4, including 813 quasars, in which the SDSS selection completeness is ≳90% for an observed quasar magnitude *i*~19 (the control sample has a median *i*~18.1 magnitude), to limit the uncertainty due to different selection completeness for BAL and non-BAL quasars. We find a BAL fraction of 15.5±1.5%, close to that found for the complete first SDSS control sample. As a further check, we consider an additional sample of (iv) 313 SDSS DR7 quasars in the redshift range 3.6-4.6, where the completeness is >90% (as for the first quasar control sample, we selected only quasars with a K band detection). At these redshifts, SDSS spectroscopy covers the 1216-1640 Å wavelength range. The BAL fraction is 19.2 ± 1.7%, again very similar to the fraction found for the first SDSS control sample. We conclude that selection completeness does not significantly affect the BAL fraction measured using Eq. 1 and the $BI_0$ threshold of 100 km/s. Therefore, throughout this study we will compare the results on the XQR-30 sample to those of the first complete SDSS control sample. The BAL fractions estimated in the XQR-30 sample and in the comparison SDSS samples are the observed ones, without corrections for selection effects. We choose to compare the observed BAL fraction at different redshift in samples with matched selection criteria, rather than trying to estimate an "intrinsic" BAL fraction (see eg. 19), because the uncertainty in this estimate is quite large, depending on both observational setups and quasar models, and can differ by orders of magnitudes in different redshift ranges.

**BAL identification.** The quasar rest-frame UV continuum emission is estimated by creating composite emission templates based on the catalogue of 11800 quasars at 2.13≲z≲3.20 from SDSS DR7 as defined above. For each XQR-30 spectrum, the template is the median of 100 non-BAL[11] quasar spectra, randomly selected from SDSS in the redshift range 2.13≲z≲3.20

(to cover the 1216-2100 Å rest wavelength range), with colors and C IV equivalent width in a range encompassing ±20% the values measured for the XQR-30 quasar. For the quasar colors we consider the F(1700Å)/F(2100Å) and F(1290Å)/F(1700Å) flux ratios, F(1700Å) and F(2100Å) being median value over 100Å and F(1290Å) being a median value over 30Å in the rest-frame. We assume that observed and intrinsic continuum emission do not differ at ~1290Å (except in the case of PSOJ065+01, for which we anchored the fit to the emission at 1330-1350Å), the bluest spectral region in which continuum emission can be probed in the XQR-30 spectra owing to the strong Lyα forest absorption at $\lambda \lesssim 1216$Å. This may represent a lower limit to the intrinsic quasar emission in the case that BAL absorption troughs affect the ~1290Å spectral region. For XQR-30 sources with redder colors, that is $F(1700Å)/F(2100Å) \lesssim 1$ and $F(1290Å)/F(1700Å) < 1$, the composite spectrum is built from 30-50 SDSS quasars that match our criteria, due to the relative rarity of red quasars in SDSS. The combined template has a spectral resolution of ~70 km/s in the C IV spectral region, which corresponds to the lower spectral resolution of the individual SDSS spectra, and is normalised to the median value of the XQR-30 quasar spectrum in the rest-frame 1650-1750 Å interval. The latter has been chosen as it is free from prominent emission lines and strong telluric absorption in the X-shooter spectra.

The normalised XQR-30 spectra are then obtained by dividing each X-shooter spectrum by the matched composite SDSS template. We systematically investigate for absorption troughs associated with C IV, Si IV, N V and Mg II ions. Our analysis is limited to blueshifted velocities up to $v_{lim} \sim 64500$km/s with respect to CIV, since in z~6 quasar spectra almost no transmitted flux is observed blueward of the Lyα emission. Given that the C IV optical depth usually dominates the Si IV depth in BAL quasars[31], we use the velocity range of the C IV BAL troughs to identify absorption associated with Si IV[27]. This implies that broad absorption features blueward of Si IV are ascribed to a low velocity Si IV BAL if a C IV BAL with similar velocity is observed in the X-shooter spectrum. This is the case of XQR-30 quasars J0923+0402, PSOJ089-015, VDESJ2211-3206, PSOJ239-07, PSOJ023-02. Otherwise, broad absorption features blueward of Si IV are due to a high-velocity ($v_{max} \gtrsim 30000$ km/s) C IV BAL. We use the same methodology to identify NV BAL features.

We identify 14 C IV BAL quasars (Table 1), of which J0923+0402, PSOJ089-15, VDESJ2211-3206, and PSOJ231-20 had been already classified as BAL by visual inspection[45,17,37,54]. Among the C IV BAL quasars, 8 also show Si IV BAL features, 3 a N V BAL and 1 a Mg II BAL. The characterisation of non-C IV BAL outflows will be presented by Bischetti et al. in preparation.

**Nuclear properties.** $M_{BH}$ and $\lambda_{Edd}$ for the XQR-30 sample have been derived by spectral modelling of the Mg II spectral region in the X-shooter data, by taking into account the quasar power-law continuum and the Balmer pseudo-continuum emission[43,17], the FeII emission[55], and one or more Gaussian components to model the Mg II emission line profile. Details of the spectral analysis, as also the individual $M_{BH}$ and $\lambda_{Edd}$ will be presented by Mazzucchelli et al. in preparation. We calculate $\lambda_{Edd} = L_{bol}/L_{Edd}$, where $L_{bol}$ is derived from the monochromatic luminosity at 3000 Å via bolometric correction[56,57]; $M_{BH}$ is based on the total full width at half maximum of the Mg II line profile[43]. The same method has been applied to the SDSS control sample quasars with Mg II measurements[12] ($z \lesssim 2.3$). In particular, the same FeII modelling and single epoch relation[58] have been used. For the remaining SDSS quasars (z>2.3) $M_{BH}$ and $\lambda_{Edd}$ are based on the C IV line[12,59]. We correct for the presence of non-virial motions affecting the C IV profile[60].

**Acknowledgments**
Based on observations collected at the European Organisation for Astronomical Research in the Southern Hemisphere under ESO large programme 1104.A-0026(A). Funding for the SDSS and SDSS-II has been provided by the Alfred P. Sloan Foundation, the Participating Institutions, the National Science Foundation, the U.S. Department of Energy, the National Aeronautics and Space Administration, the Japanese Monbukagakusho, the Max Planck Society, and the Higher Education Funding Council for England. The SDSS Web Site is http://www.sdss.org/. The SDSS is managed by the Astrophysical Research Consortium for the Participating Institutions. The Participating Institutions are the American Museum of Natural History, Astrophysical Institute Potsdam, University of Basel, University of Cambridge, Case Western Reserve University, University of Chicago, Drexel University, Fermilab, the Institute for Advanced Study, the Japan Participation Group, Johns Hopkins University, the Joint Institute for Nuclear Astrophysics, the Kavli Institute for Particle Astrophysics and Cosmology, the Korean Scientist Group, the Chinese Academy of Sciences (LAMOST), Los Alamos National Laboratory, the Max-Planck-Institute for Astronomy (MPIA), the Max-Planck-Institute for Astrophysics (MPA), New Mexico State University, Ohio State University, University of Pittsburgh, University of Portsmouth, Princeton University, the United States Naval Observatory, and the University of Washington.

M.B., C.F. and F.F. acknowledge support from PRIN MIUR project "Black Hole winds and the Baryon Life Cycle of Galaxies: the stone-guest at the galaxy evolution supper", contract #2017PH3WAT. RLD is supported by a Gruber Foundation Fellowship grant. GB was supported by NSF grant AST-1751404. SEIB acknowledges funding from the European Research Council (ERC) under the European Union's Horizon 2020 research and innovation programme (grant agreement No. 740246 "Cosmic Gas". This research was conducted by the Australian Research Council Centre of Excellence for All Sky Astrophysics in 3 Dimensions (ASTRO 3D), through project number CE170100013. This paper includes data gathered with the 6.5m Magellan Telescopes located at Las Campanas Observatory, Chile. Based on observations collected at the European Southern Observatory under ESO programme 0102.A-0233(A).

**Materials & Correspondence**
Correspondence should be addressed to Manuela Bischetti (manuela.bischetti@inaf.it). Datasets generated during and/or analysed for this study are available from the corresponding author on reasonable request.

**Supplemental material**

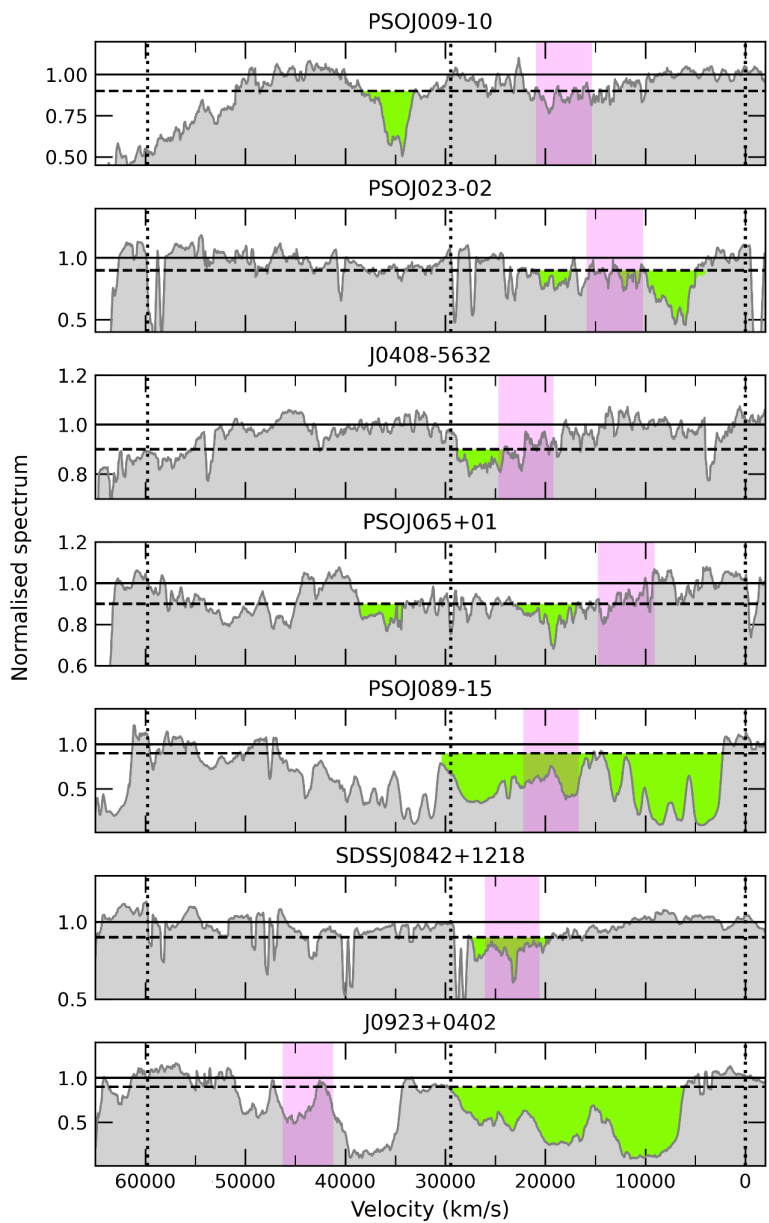

**Figure S1.** Normalised spectra of BAL quasars in the XQR-30 sample, smoothed to 500 km s−1. The velocity axis in each panel is relative to the rest-frame wavelength of C IV. Vertical dotted lines at v=0, 29000,60000 km/s indicate the velocity associated with C IV, Si IV and N V emission lines, respectively. The solid(dashed) horizontal line represents a flux level of 1.0(0.9). C IV BALs, corresponding to a flux level <0.9 (Eq. 1), are highlighted as green shaded areas. We note that the C IV optical depth typically dominates that of Si IV in BAL quasars. This implies that e.g. the BAL feature at v~-45000 in PSOJ009-10 cannot be ascribed to a low velocity Si IV BAL because no low velocity C IV BAL with similar velocity is observed in the X-shooter spectrum. The shaded magenta areas indicate the overlapping spectral region between the X-shooter VIS and NIR arms.

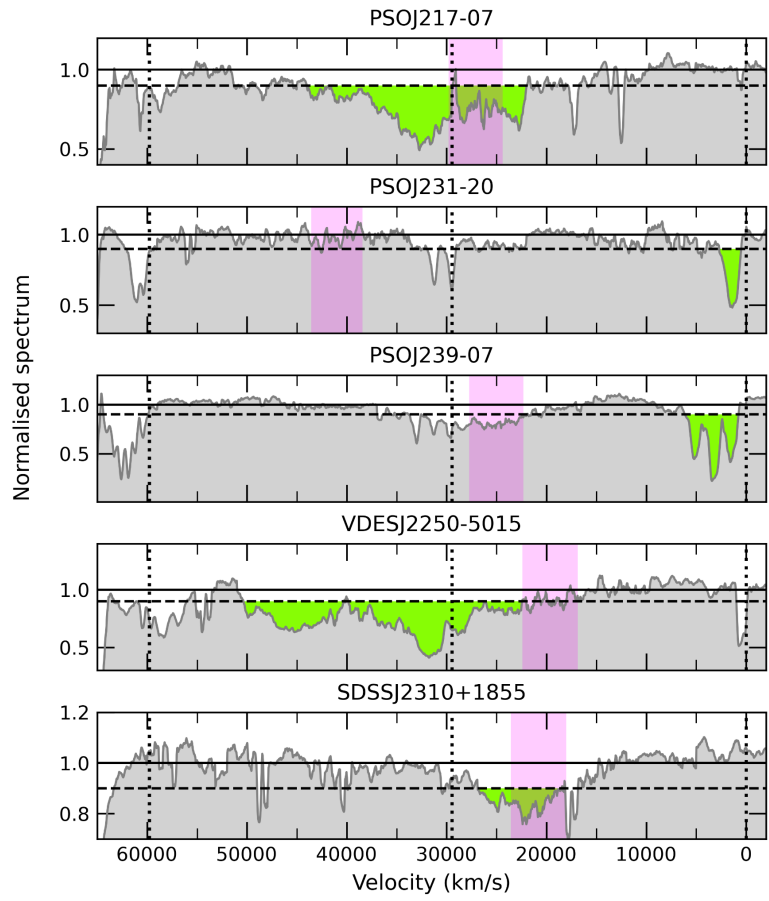

**Figure S2.** Same as Figure S1.